\documentclass[letterpaper]{article} % DO NOT CHANGE THIS
\usepackage{aaai25}  % DO NOT CHANGE THIS
\usepackage{times}  % DO NOT CHANGE THIS
\usepackage{helvet}  % DO NOT CHANGE THIS
\usepackage{courier}  % DO NOT CHANGE THIS
\usepackage[hyphens]{url}  % DO NOT CHANGE THIS
\usepackage{graphicx} % DO NOT CHANGE THIS
\usepackage{natbib}  % DO NOT CHANGE THIS AND DO NOT ADD ANY OPTIONS TO IT
\usepackage{caption} % DO NOT CHANGE THIS AND DO NOT ADD ANY OPTIONS TO IT
\usepackage{algorithm}
\usepackage{algorithmic}
\usepackage{amsmath,amssymb,amsfonts}
\usepackage{multirow}
\usepackage{booktabs}
\usepackage{bm}
\usepackage{tabularray}
\usepackage{newfloat}
\usepackage{listings}
\DeclareCaptionStyle{ruled}{labelfont=normalfont,labelsep=colon,strut=off} % DO NOT CHANGE THIS
\floatstyle{ruled}
\newfloat{listing}{tb}{lst}{}
\floatname{listing}{Listing}
\title{FactorGCL: A Hypergraph-Based Factor Model with Temporal Residual Contrastive Learning for Stock Returns Prediction}

\author{
    Yitong Duan,
    Weiran Wang,
    Jian Li
}
\affiliations{
    Tsinghua University, Beijing, China\\
    \{dyt19, wang-wr21\}@mails.tsinghua.edu.cn, lijian83@mail.tsinghua.edu.cn
}

\begin{document}

\maketitle

\begin{abstract}
    As a fundamental method in economics and finance, the factor model has been extensively utilized in quantitative investment. In recent years, there has been a paradigm shift from traditional linear models with expert-designed factors to more flexible nonlinear machine learning-based models with data-driven factors, aiming to enhance the effectiveness of these factor models. However, due to the low signal-to-noise ratio in market data, mining effective factors in data-driven models remains challenging. In this work, we propose a hypergraph-based factor model with temporal residual contrastive learning (FactorGCL) that employs a hypergraph structure to better capture high-order nonlinear relationships among stock returns and factors. To mine hidden factors that supplement human-designed prior factors for predicting stock returns, we design a cascading residual hypergraph architecture, in which the hidden factors are extracted from the residual information after removing the influence of prior factors. Additionally, we propose a temporal residual contrastive learning method to guide the extraction of effective and comprehensive hidden factors by contrasting stock-specific residual information over different time periods. Our extensive experiments on real stock market data demonstrate that FactorGCL not only outperforms existing state-of-the-art methods but also mines effective hidden factors for predicting stock returns.
\end{abstract}

\section{Introduction}
In the domain of stock investment, the factor model has long been a cornerstone for explaining and predicting stock returns, which employs specific variables, known as factors, to elucidate fluctuations in stock prices \cite{daniel2020short, fama2021multifactor}. This method, prevalent in both academia and industry, has demonstrated a strong ability to explain and predict stock returns. Consequently, establishing an effective factor model is of paramount importance in stock investment.

\begin{figure}[t]
    \begin{centering}
        \includegraphics[width=0.95\columnwidth]{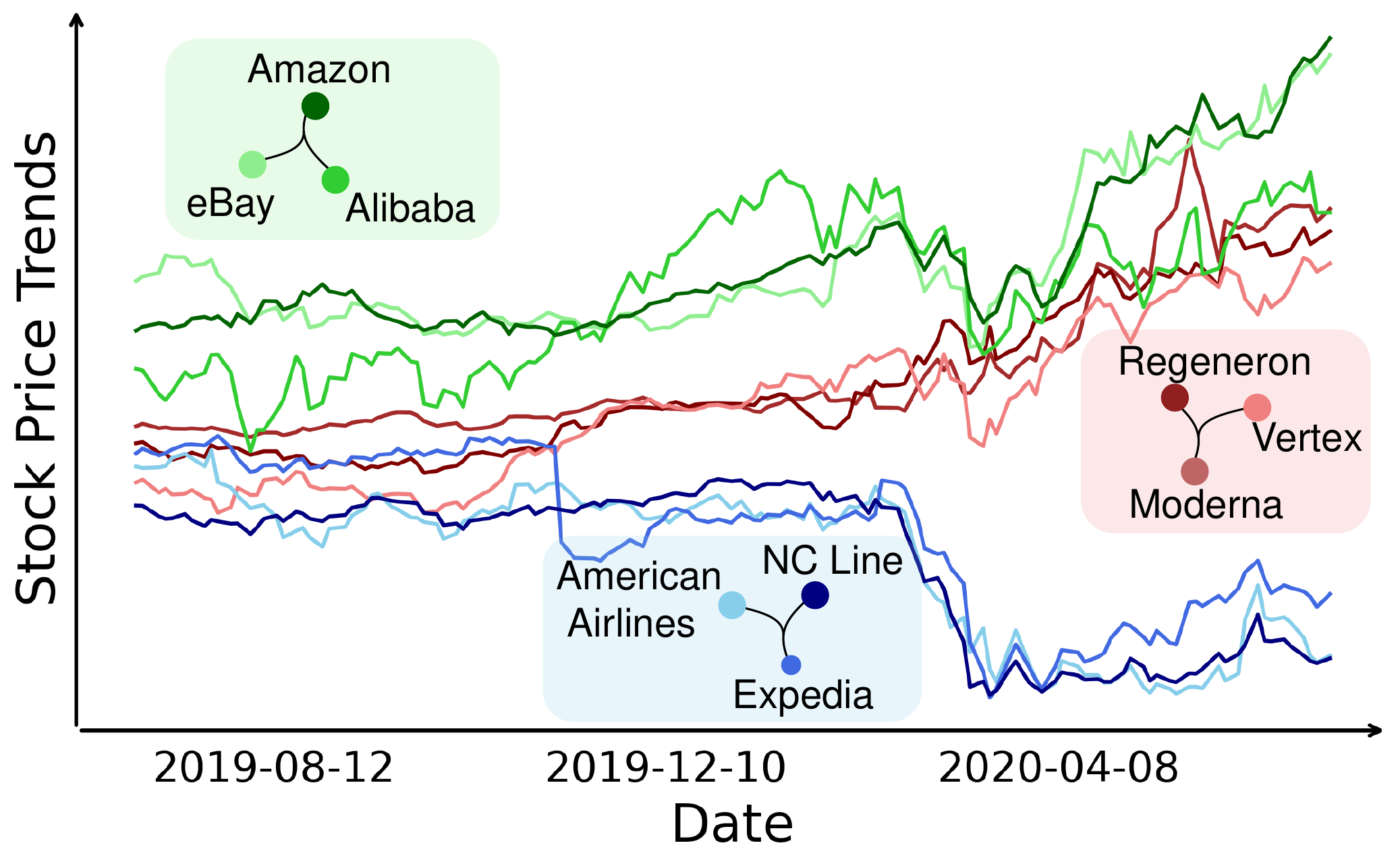}
        \caption{Stock price trends vary across different sectors and industries. During the COVID-19 pandemic, the stock price trends of the electronic consumer and medical industries exhibited high correlations, which are insufficiently explained using human-designed industry factors.}
        \label{fig:stock_trend}
    \end{centering}
\end{figure}

The factor model explains stock returns by utilizing various factors, including fundamental, technical, and macroeconomic indicators. Specifically, in a factor model, stocks are described by factors and their corresponding factor exposures, which represent the impact of factors on stocks. In traditional factor models, these factors are designed based on expert practical experience. For instance, the well-known Fama-French model \cite{eugene1992cross} employs three manually designed factors: \textit{market}, \textit{size}, and \textit{value}. However, these human-designed factors, while effective, are limited in number and may not sufficiently explain stock returns. For example, as illustrated in Figure~\ref{fig:stock_trend}, stock price trends across different industries exhibited high correlations that could not be adequately explained by industry-specific factors based on prior human experience. Moreover, most existing factor models are linear, explaining stock returns through a linear combination of factors weighted by factor exposures. However, recent studies \cite{levin1995stock, almeida2023nonlinear, bansal2004risks, he2013intermediary} have identified complex nonlinear relationships between factors and stock returns in real markets. This discrepancy highlights the limitations of linear factor models in capturing actual market behavior. Consequently, a core issue in current factor model research is how to mine more effective factors that are applicable to real market behavior.

Recent advancements in machine learning (ML) have introduced a new perspective for factor model research \cite{kelly2019characteristics, uddin2020latent, gu2021autoencoder}. ML-based approaches can learn complex and nonlinear market patterns in a data-driven manner and construct more market-adaptive models \cite{li2024alphafin, zou2022astock, cui2023temporal}. However, the low signal-to-noise ratio in stock market data may complicate the learning process for ML-based factor models. Most current ML-based methods extract factors from market data without effectively leveraging prior human experience, which may lead to overfitting of the extracted factors to market noise rather than capturing effective patterns. This limitation presents a significant obstacle to the application of ML in factor models.

\begin{figure}[t]
    \begin{centering}
        \includegraphics[width=0.75\columnwidth]{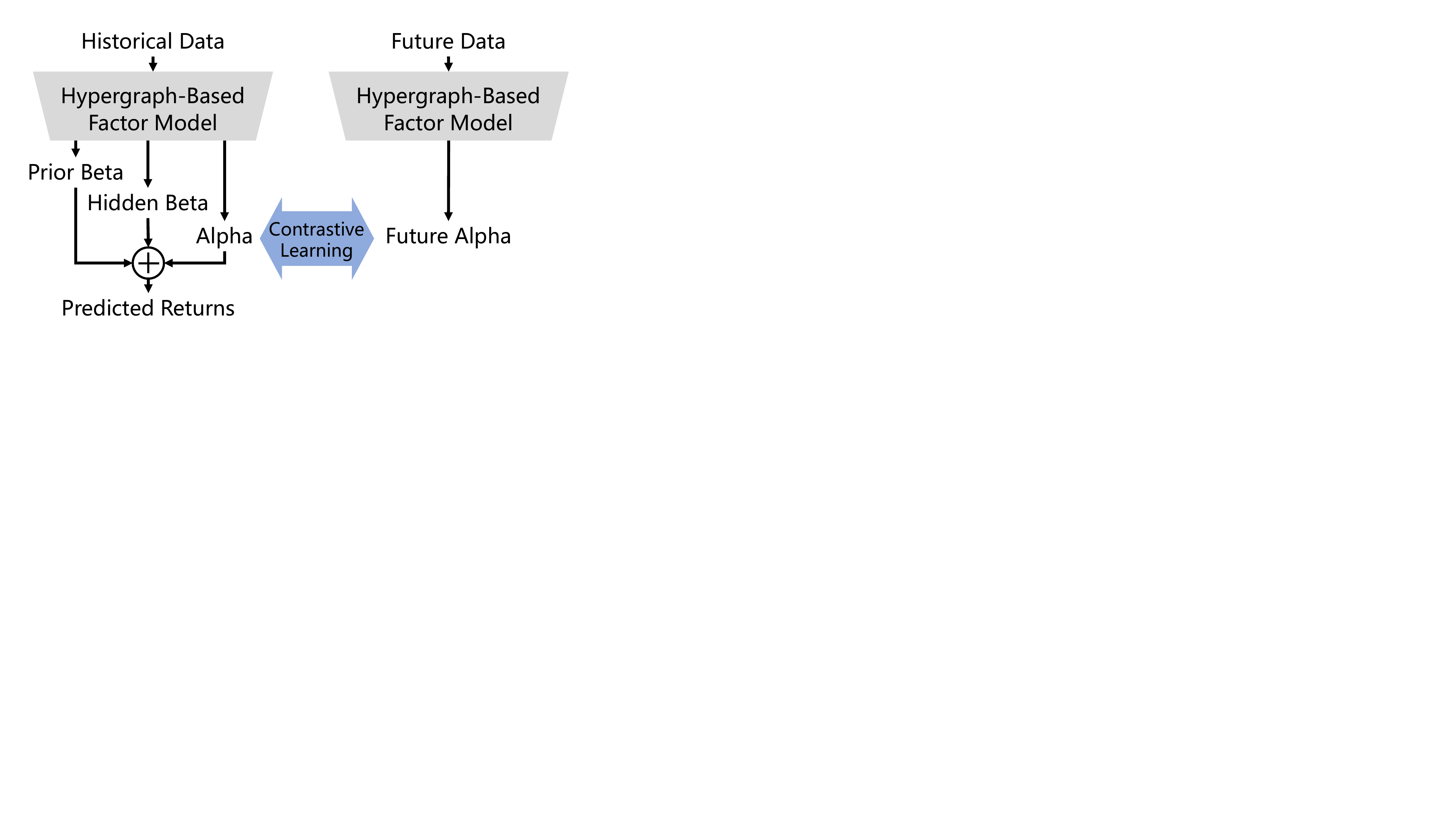}
        \caption{Brief illustration of FactorGCL.}
        \label{fig:overview}
    \end{centering}
\end{figure}

To address this obstacle, we propose a hypergraph-based factor model with temporal residual contrastive learning (FactorGCL) that supplements human-designed prior factors with data-driven hidden factors, thereby enhancing the effectiveness of factor models in predicting stock returns. Our model utilizes a hypergraph, a generalized graph structure, to better capture high-order relationships among stock returns and factors. Specifically, stocks are treated as nodes in the hypergraph, factors are represented as hyperedges, and the mining of hidden factors is framed as a hyperedge generation task. As shown in Figure~\ref{fig:overview}, FactorGCL employs a cascading residual hypergraph architecture where stock returns are decomposed into three components: prior beta, hidden beta, and individual alpha. Each component is extracted from the residuals after removing the influence of the previous component. Additionally, we propose a temporal residual contrastive learning method to guide the model's learning process by contrasting individual stock residuals across different time periods.

In summary, the contributions of our work are as follows:
\begin{itemize}

    \item We propose FactorGCL, a novel factor model that utilizes a hypergraph structure to capture high-order nonlinear relationships among stock returns and factors. It employs a cascading residual hypergraph architecture to mine hidden factors, supplementing human-designed prior factors, thereby enhancing the prediction of stock returns.

    \item We design a self-supervised learning method called temporal residual contrastive learning. This method enhances the model's ability to extract effective and comprehensive hidden factors by contrasting stock-specific residuals across different time periods to better guide the mining of hidden factors.

    \item We conduct extensive experiments on real stock market data. The results demonstrate that our method not only surpasses existing state-of-the-art baselines in stock trend prediction but also mines effective hidden factors for predicting stock returns.

\end{itemize}

\section{Related Work}

\subsection{Factor Model}
Factor models are widely utilized in stock investments. The original factor model, the capital asset pricing model (CAPM) \cite{treynor1961toward, sharpe1964capital, lintner1975valuation}, attributes differences in stock returns to varying exposures to a single market factor. Later, in a seminal work \cite{eugene1992cross}, it was observed that firm value and size also contribute to explaining expected stock returns, and proposed the Fama-French three-factor model.
With the advancement of machine learning, some machine learning-based factor models have emerged. \cite{levin1995stock} proposed a nonlinear factor model based on neural networks to model possible interactions between different factors. \cite{gu2021autoencoder} proposed a latent dynamic factor model using a conditional autoencoder network to capture non-linearity in return dynamics, demonstrating that the nonlinear factor model outperforms other leading linear methods. Furthermore, \cite{duan2022factorvae} introduced a probabilistic factor model based on variational autoencoders to better extract effective factors from the market data with high noise levels.

\subsection{Hypergraph Neural Network}
Hypergraphs have proven to be an efficient approach for modeling high-order correlations among data. \cite{zhou2006learning} first introduced hypergraph learning as a propagation process on hypergraph structures. \cite{feng2019hypergraph} further advanced this concept by developing the hypergraph convolutional neural network using deep learning methods for data representation learning. Hypergraphs have also been widely applied in the field of stock return prediction\cite{li2022hypergraph, han2023stock, su2024attention}, \cite{sawhney2020spatiotemporal} initially applied the hypergraph neural network to learn stock price evolution based on stock relationships. Subsequently, \cite{sawhney2021stock} improved hypergraph neural network for stock trend prediction by incorporating ranking loss. \cite{xu2021hist} designed a concept-oriented graph framework to mine hidden concepts for stock trend forecasting. Additionally, \cite{xia2024ci} developed a dynamic hypergraph for stock selection problem using a transformer-based pretraining mechanism.

\subsection{Contrastive Learning}
This work is also related to contrastive learning , a promising class of self-supervised methods that leverage the semantic dependencies of sample pairs to capture the essence of data \cite{chen2021exploring, lin2022prototypical, li2020prototypical, caron2020unsupervised}. Contrastive learning has been widely used in various applications. For instance, \cite{oord2018representation} introduced Contrastive Predictive Coding to learn useful representations for predicting future data. \cite{liu2022towards} proposed a unsupervised deep graph structure learning method based on contrastive learning. In the financial domain, \cite{hou2021stock} introduced a contrastive learning method for multi-granularity stock data and used it as a regularization term to improve stock trend prediction tasks.

\section{Preliminaries}
% In this section, we first introduce the basic concepts of hypergraphs and hypergraph convolutional neural networks. We then outline the fundamental principles of contrastive learning.

In this section, we first introduce the basic concepts of hypergraph convolutional neural networks, and then formally describe the research problem.

\subsection{Hypergraph Convolutional Neural Network}

A hypergraph generalizes a graph by allowing an edge, termed a hyperedge, to connect two or more nodes. This structure enables the hypergraph to capture the group-wise correlations beyond pair-wise connections. Formally, a hypergraph is defined as \( \mathcal{G} = (\mathcal{V}, \mathcal{E}, W) \), where \( V \) and \( E \) denote the sets of vertices and hyperedges, respectively, and \( W \) is a diagonal matrix assigning weights to the hyperedges. The pair \((\mathcal{V}, \mathcal{E})\) in a hypergraph can be represented by an incidence matrix \( H \in \mathbb{R}^{|\mathcal{V}| \times |\mathcal{E}|} \), where \( H^{(i, j)} \) indicates the connection between the \(i\)-th vertex \( \mathcal{V}^{(i)} \) and the \(j\)-th hyperedge \( \mathcal{E}^{(j)} \), defined as:
\(
H^{(i, j)} =
\begin{cases}
    1, & \text{if } \mathcal{V}^{(i)} \in \mathcal{E}^{(j)}    \\
    0, & \text{if } \mathcal{V}^{(i)} \notin \mathcal{E}^{(j)}
\end{cases}
\)

The hypergraph convolutional neural network (HyperGCN) \cite{feng2019hypergraph} extends graph convolutional networks to hypergraphs, enabling the capture of high-order relationships inherent in hypergraph structures. The core of HyperGCN involves a message propagation rule that updates node features by aggregating information from their connected hyperedges, which are influenced by all nodes connected by those hyperedges. Formally, the node features at the \((l+1)\)-th layer of HyperGCN, \(e^{(l+1)} \in \mathbb{R}^{|\mathcal{V}| \times d_{l+1}}\), are computed using the formula:
\begin{equation}
    e^{(l+1)} = \sigma(D_{n}^{-1/2} H W D_{e}^{-1} H^T D_{n}^{-1/2} e^{(l)} w)
\end{equation}
where \( \sigma \) is a non-linear activation function, \(D_{n} \in \mathbb{R}^{|\mathcal{V}| \times |\mathcal{V}|}\) and \(D_{e} \in \mathbb{R}^{|\mathcal{E}| \times |\mathcal{E}|}\) are diagonal matrices representing node degrees and hyperedge degrees, respectively. \(W \in \mathbb{R}^{|\mathcal{E}| \times |\mathcal{E}|}\) is a diagonal matrix representing hyperedge weights, \( w \in \mathbb{R}^{d_l \times d_{l+1}}\) is a learnable weight matrix, with \(d_l\) representing the dimensions of node features at the \(l\)-th layer.

\subsection{Problem Formulation}
% \(S = \{s_1, s_2, \ldots, s_N\}\) \(F = \{f_1, f_2, \ldots, f_K\}\)
Given $N$ stocks in cross-section of the stock market, and a set of $K$ factors, the traditional linear factor model calculates expected stock returns as a linear combination of factors weighted by factor exposures:
\begin{equation}
    \begin{aligned}
        y_{t} = \sum_{k=1}^{K}\beta_{t}^{(k)} z_{t}^{(k)} + \alpha_{t}
    \end{aligned}
\end{equation}
where \(y_{t} = \frac{\text{price}_{t+\Delta t} - \text{price}_{t}}{\text{price}_{t}} \in \mathbb{R}^{N}\) denotes the future returns of \(N\) stocks at trading day \(t\), \(\beta_{t} \in \mathbb{R}^{N \times K}\) is the factor exposure matrix of stocks at trading day \(t\), \(\beta_{t}^{(k)} \in \mathbb{R}^{N}\) represents the \(k\)-th factor exposure of stocks at trading day \(t\), \(z_{t} \in \mathbb{R}^{K}\) is the vector of \(K\) factor returns, and \(\alpha_{t} \in \mathbb{R}^{N}\) denotes the idiosyncratic returns of the stocks.

\cite{levin1995stock} extends the linear factor model to a nonlinear version, where the stock returns are calculated using a nonlinear function $h$ of factor returns and factor exposures:
\begin{equation}
    \begin{aligned}
        {y_{t}} = h(\beta_{t}, z_{t}) + \alpha_{t}
    \end{aligned}
\end{equation}

Typically, factor returns and stock individual returns are estimated using factor exposures and historical stock returns. For example, in a linear factor model, factor returns \( z \) are estimated using the slopes from the linear regression on historical data, while individual returns \( \alpha \) are estimated using the residuals. Therefore, we set $z_t = f_{z}(\beta_t, x_t)$ and $\alpha_t = f_{\alpha}(\beta_t, x_t)$, where $x_t \in \mathbb{R}^{N\times T \times D}$ represents the historical data of stocks at trading day $t$, with \( T \) being the length of historical data and \( D \) being the feature dimension of each stock's data.

Finally, the task in this work is to learn a nonlinear factor model based on given factor exposures $\beta_{t}$ and historical stock market data $x_{t}$, for predicting future stock returns.

\begin{equation}
    \begin{aligned}
        {\hat{y}_{t}}=h(\beta_{t}, z_t) + \alpha_t =f_{\beta}(\beta_{t}, x_t) + f_{\alpha}(\beta_{t}, x_t)
    \end{aligned}
\end{equation}

In the following sections, we simplify our notation by omitting the time subscript \(t\). Unless otherwise specified, all references to \(w\) and \(b\) pertain to the weights and biases of linear layers, respectively, and will not be further elaborated.

\section{Methodology}

% This section details the design of FactorGCL. We first formally describe the research problem, and then introduce the cascading residual hypergraph architecture of our model, which extracts prior beta, hidden beta, and individual alpha components to predict stock returns. Next, we present a temporal residual contrastive learning method to guide the model in extracting effective and comprehensive hidden factors.

This section introduces the design of FactorGCL. We first design a cascading residual hypergraph architecture for our model, which extracts prior beta, hidden beta, and individual alpha components to predict stock returns. Next, we propose a temporal residual contrastive learning method to guide the model in extracting effective and comprehensive hidden factors.

\begin{figure}[t]
    \begin{centering}
        \includegraphics[width=0.99\columnwidth]{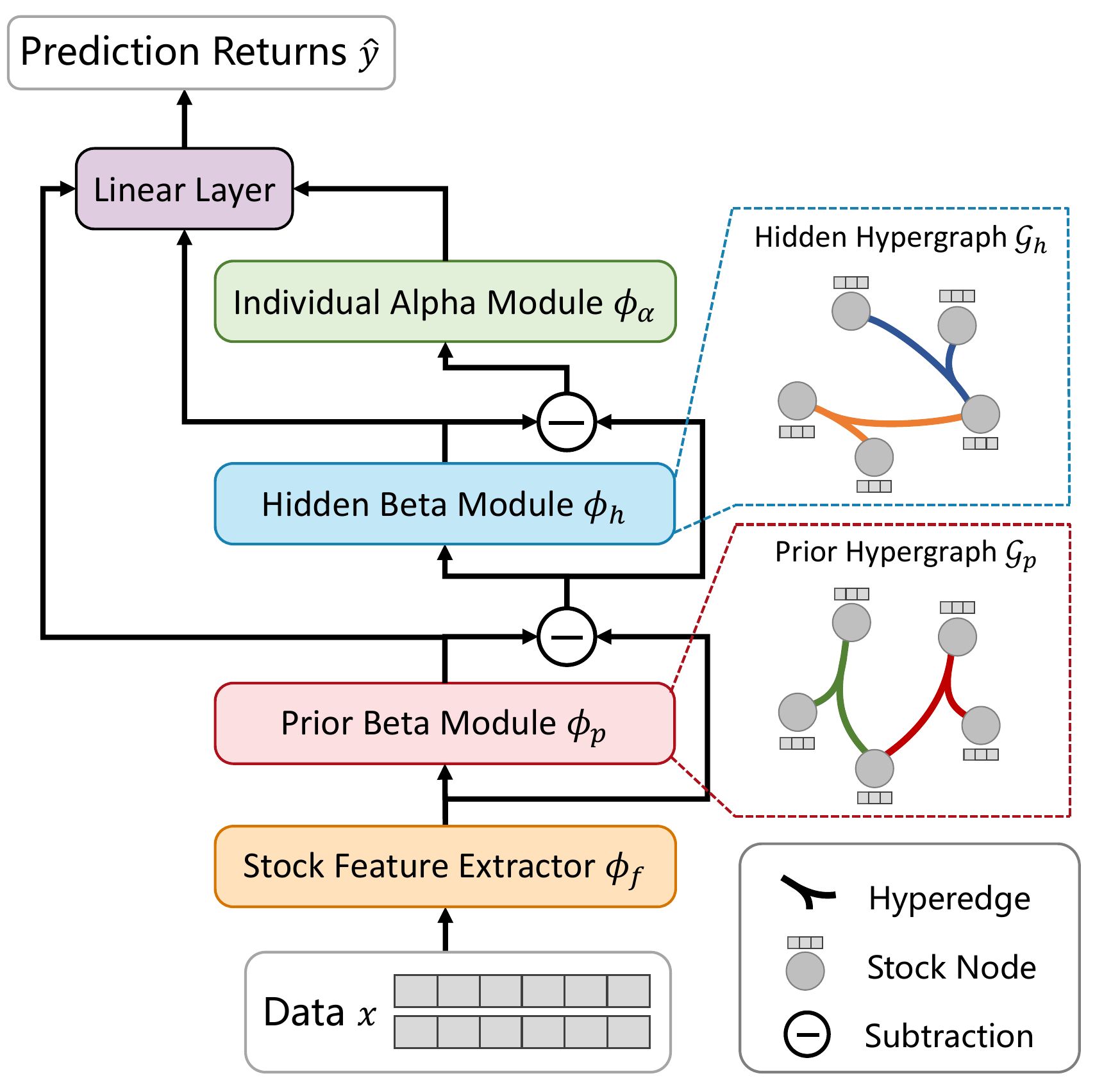}
        \caption{Overview of the cascading residual hypergraph architecture in FactorGCL. Stock returns are decomposed into prior beta, hidden beta, and individual alpha components. Each component is extracted from the residuals after removing the influence of the preceding component.}
        \label{fig:residual}
    \end{centering}
\end{figure}

\subsection{Cascading Residual Hypergraph Architecture}

As previously mentioned, we utilize hypergraphs to construct a nonlinear factor model. Inspired by \cite{xu2021hist}, we design a cascading residual hypergraph architecture to better extract hidden factors. As illustrated in Figure~\ref{fig:residual}, we decompose the predicted stock returns into three components: prior beta, hidden beta, and individual alpha. Specifically, we first extract stock features from the raw data and then use the prior beta module to extract the representations of prior factors. Next, we mine hidden factors from the residuals after removing the prior beta information using the hidden beta module. Finally, we extract individual alpha from the residuals after removing both prior and hidden beta information. The final prediction of our model is obtained by summing these three components. The detailed design of our architecture is as follows.

\subsubsection{Feature Extractor}

Given the raw sequential market data \(x \in \mathbb{R}^{N \times T \times D}\), the feature extractor \(\phi_{\text{feat}}\) encodes the stock temporal feature \(e_s \in \mathbb{R}^{N \times H}\) to capture rich temporal information. This process is defined as \(e_s = \phi_{\text{feat}}(x)\), where \(H\) represents the dimension of the feature embeddings. To capture long-term dependencies in sequential data, we utilize a gated recurrent unit with a batch normalization as the feature extractor, using the hidden state at the last time step as the stock feature embeddings.

\subsubsection{Prior Beta Module}
\begin{figure}[t]
    \begin{centering}
        \includegraphics[width=0.99\columnwidth]{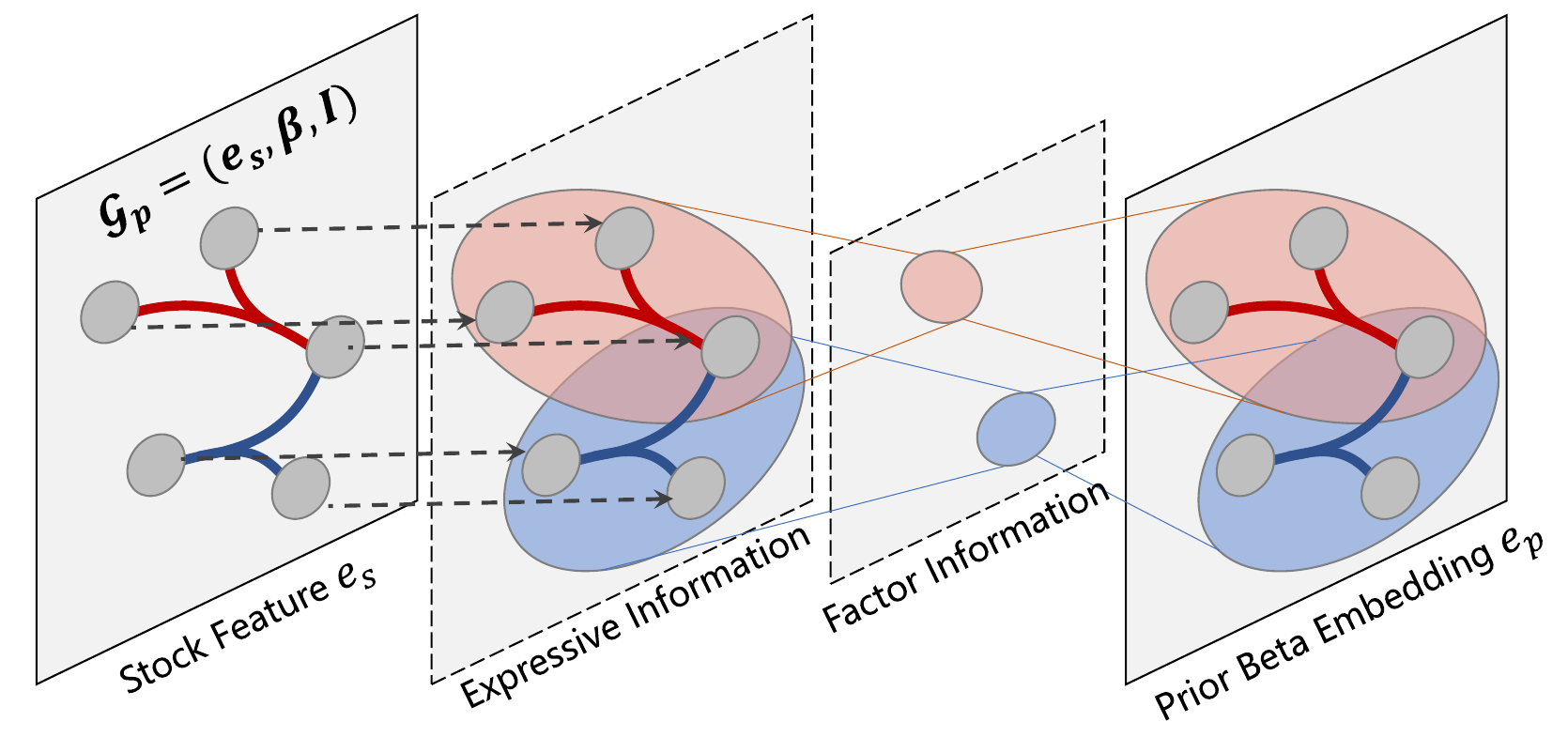}
        \caption{Illustration of the information propagation process in the HyperGCN.
            The HyperGCN can model the nonlinear influence of factors on stocks by aggregating information from stock nodes connected by the same hyperedge.}
        \label{fig:prior}
    \end{centering}
\end{figure}

To leverage expert knowledge from given $K$ prior factors \(\beta \in \mathbb{R}^{N \times K}\), we employ a hypergraph convolutional neural network (HyperGCN) to model the nonlinear relationships among stock returns and these factors. Specifically, we represent stocks as nodes in the hypergraph and factors as hyperedges. Stocks exposed to the same factor are connected by the same hyperedge, with the incidence matrix representing factor exposures. We posit that the information propagation mechanism in the HyperGCN effectively captures the nonlinear influence of factors on stocks. This process, illustrated in Figure~\ref{fig:prior}, comprises the following steps:

\begin{itemize}
    \item \textbf{Message extraction}: Applies a transformation matrix to the each node features to extract expressive information.
    \item \textbf{Message aggregation}: Aggregates the information from stock nodes connected by the same hyperedge, representing the shared information of the corresponding factor.
    \item \textbf{Message sharing}: Integrates the node embeddings with the shared factor information as the influence of factors on stocks.
\end{itemize}

Formally, given the stock feature embeddings \( e_s \) output by the feature extractor, we build a hypergraph \(\mathcal{G}_{p}\) with node features \(e_s \) and incidence matrix \(\beta \). We then calculate the prior beta embeddings by applying the HyperGCN to the hypergraph \(\mathcal{G}_{p}\), expressed as:

\begin{equation}
    \begin{aligned}
        e_{p} & = \phi_{\text{prior}}(e_s, \beta)                                        \\
              & = \sigma(D_{n}^{-1/2} \beta W D_{e}^{-1} \beta^T D_{n}^{-1/2} e_s w_{p})
    \end{aligned}
\end{equation}
where \( \sigma \) is the LeakyReLU activation function, \( D_{n} \) and \( D_{e} \) are the degree matrices of the nodes and hyperedges, respectively, and \( W = I \) is the identity matrix. The resulting prior beta embeddings \( e_{p} \in \mathbb{R}^{N \times H} \) represent the influence of prior factors on stocks.

\subsubsection{Hidden Beta Module}

As previously mentioned, factors based on human prior knowledge may not adequately capture stock returns. To address this, we designed a hidden beta module to extract hidden factors that supplement these prior factors. Specifically, we regard the extraction of hidden factors as a hyperedge generation task: after removing the prior factor information, the hidden beta module generates new hyperedges from the residual embeddings and constructs a new hypergraph to model the nonlinear influence of these hidden factors on stocks.

Formally, we first calculate the residual embeddings by subtracting the prior factor embeddings from the stock feature embeddings, i.e., \( e_r = e_s - e_p \). Next, we construct \( M \) learnable vectors \( \{c^{(i)}\}_{i=1}^{M} \), where \( c^{(i)} \in \mathbb{R}^{H} \), referred to as hidden factor prototypes. Hidden factors are mined by calculating the similarity between the residual embeddings and the hidden factor prototypes:$
    \beta_{h}^{(i, j)} = \text{Sigmoid}(e_r^{(i)} \cdot {c^{(j)}}^{T})
$
where \( \beta_{h} \in \mathbb{R}^{N \times M} \) is the hidden factor exposure matrix.

Similar to the prior beta module, we construct a hypergraph \(\mathcal{G}_{h}\) with node features \(e_r\) and incidence matrix \(\beta_{h}\), and then extract the influence of hidden factors on stocks by applying the HyperGCN to the hypergraph \(\mathcal{G}_{h}\):

\begin{equation}
    \begin{aligned}
        e_{h}  = \phi_{\text{hidden}}(e_r, \beta_{h})  = \text{HyperGCN}(e_r, \beta_{h})
    \end{aligned}
\end{equation}
where \( e_{h} \in \mathbb{R}^{N \times H} \) is the hidden beta embeddings. Note that we generate "soft" hyperedges $\beta_{h}$ in the hidden beta module, with values ranging between [0, 1], enhancing the flexibility of the hidden factors.

\subsubsection{Individual Alpha Module}

In addition to the influence of prior and hidden factors, the idiosyncratic information of the stock itself, or alpha, also significantly impacts stock returns. The individual alpha module handles the residual embeddings after removing the prior and hidden factor embeddings to capture the stock-specific information. We calculate the individual alpha embeddings \( e_{\alpha} \in \mathbb{R}^{N \times H} \) by applying a linear layer with a LeakyReLU activation function:
\begin{equation}
    \begin{aligned}
        e_{\alpha} = \text{LeakyReLU}(w_{\alpha}(e_{s} - e_{p} - e_{h}) + b_{\alpha})
    \end{aligned}
\end{equation}

\subsubsection{Prediction}
We obtain the model's prediction by performing a linear mapping on the embeddings output by the prior beta module, hidden beta module, and individual alpha module, and then summing them up. Additionally, we design a multi-label prediction that requires the model to predict stock returns over multiple forward periods. This approach aims to enhance the robustness and reliability of our model by ensuring its predictive power extends across different future time frames.

\begin{equation}
    \begin{aligned}
        \hat{y}^{(l)} = w_{o1}^{(l)} e_{p} + w_{o2}^{(l)} e_{h} + w_{o3}^{(l)} e_{\alpha} + b_{\text{o}}^{(l)}
    \end{aligned}
\end{equation}
where \( \hat{y}^{(l)} \in \mathbb{R}^{N} \) represents the predicted stock returns of the \( l \)-th forward prediction period.

\subsection{Temporal Residual Contrastive Learning}

As previously mentioned, data-driven factor models face the challenge of a low signal-to-noise ratio in market data. Specifically, the hidden factors mined through such models encounter two main issues:

\begin{itemize}
    \item \textbf{Effectiveness}: The hidden factors extracted from historical data should remain consistently effective in the future. However, factors extracted by data-driven approaches are prone to overfitting market noise, thus lacking effectiveness in predicting future stock returns.
    \item \textbf{Comprehensiveness}: The hidden factors should supplement prior factors to provide a comprehensive description of stock returns. Nonetheless, market noise complicates factor mining, making the model more likely to extract simplistic factors while neglecting others, thereby failing to adequately represent stock returns.

\end{itemize}

\begin{figure}[t]
    \begin{centering}
        \includegraphics[width=0.9\columnwidth]{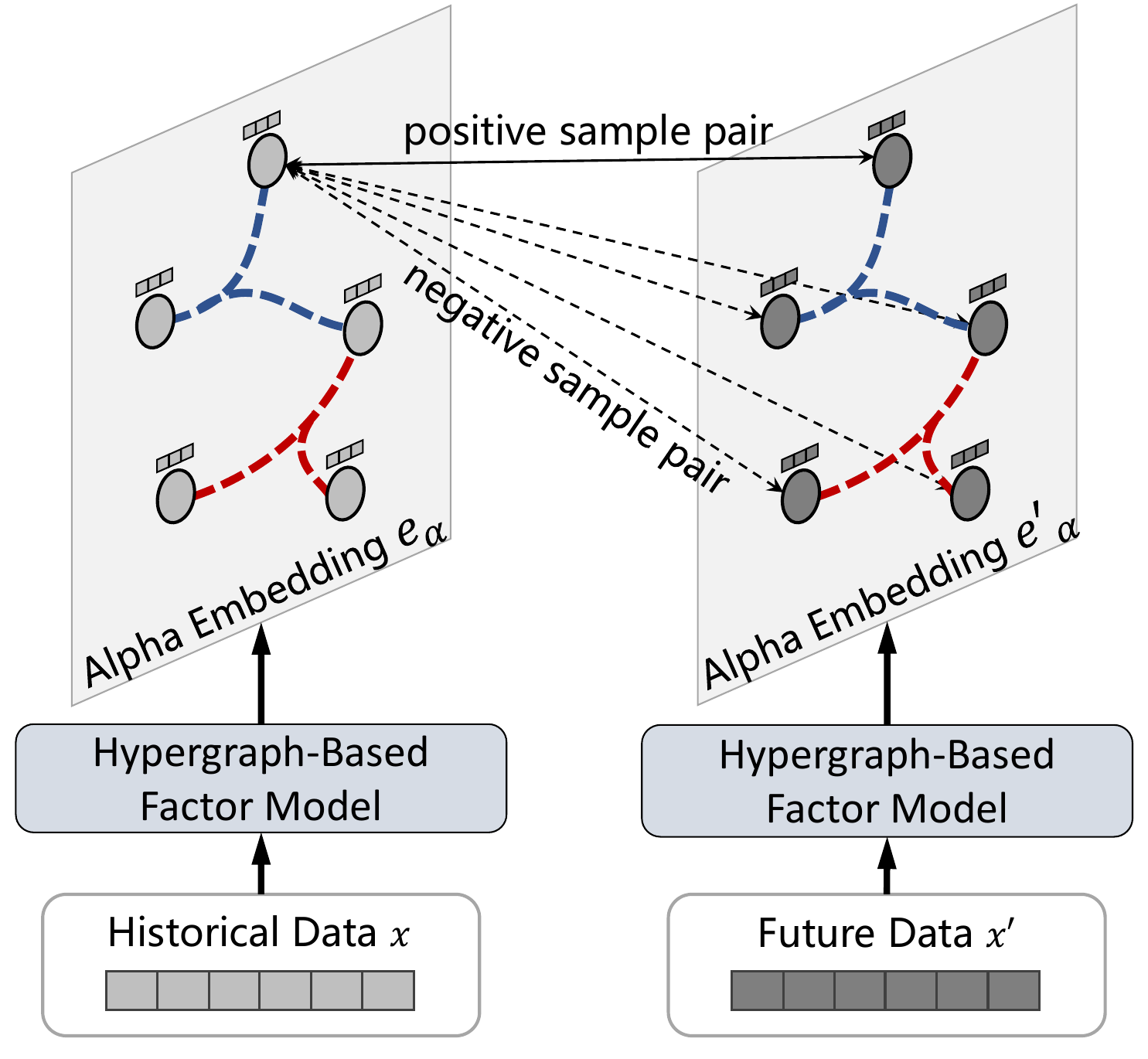}
        \caption{Illustration of the temporal residual contrastive learning method. The model contrasts the past and future alpha embeddings of the same stock as positive pairs, and the embeddings of different stocks as negative pairs.}
        \label{fig:contrastive_learning}
    \end{centering}
\end{figure}

To address these issues, we have developed a self-supervised contrastive learning method for FactorGCL, termed temporal residual contrastive learning. The motivation behind this method is as follows: for a factor model with effective and comprehensive factors, after removing all factor information, the residual, represented by the alpha embeddings \(e_{\alpha}\), should contain only idiosyncratic information unique to each stock, independent of other stocks. Based on this intuition, we draw inspiration from \cite{oord2018representation} and design a cross-temporal contrastive learning approach at the stock node level. As illustrated in the Figure~\ref{fig:contrastive_learning}, given the same prior and hidden factors, we use our model to calculate alpha embeddings based on both past and future market data. We then treat the past and future alpha embeddings of the same stock as positive pairs, and the embeddings of different stocks as negative pairs. By training the model with a cascading residual hypergraph architecture to extract temporally consistent alpha embeddings through a contrastive learning objective function, our model can be guided to mine hidden factors that are both effective and comprehensive.

Formally, given future data \( x' \in \mathbb{R}^{N \times T' \times D} \), with \( T' \) being the length of future data,
we use the prior factor exposure \( \beta \) and hidden factor exposure \( \beta_{h} \) extracted from historical data to calculate the future alpha embedding \(e_{\alpha}'\):

\begin{gather}
    e'_{s}      = \phi'_{\text{feat}}(x') \\
    e_{\alpha}' = e'_{s} - \phi'_{\text{prior}}(e'_{s}, \beta) - \phi'_{\text{hidden}}(e'_{r}, \beta_{h})
\end{gather}
where \( \phi'_{\text{feat}} \), \( \phi'_{\text{prior}} \), and \( \phi'_{\text{hidden}} \) represent the feature extractor, prior beta module, and hidden beta module for the future data, respectively. Note that \( \phi'_{\text{prior}} \) and \( \phi'_{\text{hidden}} \)  share parameters with \( \phi_{\text{prior}} \) and \( \phi_{\text{hidden}} \), respectively.

In this context, we employ the InfoNCE loss function from \cite{oord2018representation} as the contrastive learning loss function, formulated as follows:
\begin{equation}\scriptsize
    \mathcal{L}_{\text{CL}}=-\frac{1}{N} \sum_{i=1}^{N} \log \frac{\exp\left(\text{sim}(p(e_{\alpha}^{(i)}), p({e'}_{\alpha}^{(i)})) / \tau\right)}{\sum_{j=1}^{N} \exp\left(\text{sim}(p(e_{\alpha}^{(i)}), p({e'}_{\alpha}^{(j)})) / \tau\right)}
\end{equation}
where \( p(x) = \) is a 2-layer MLP with LeakyReLU activation functions, and \( \text{sim}(x, y) \) represents the cosine similarity function, \( \tau \) is the temperature parameter.

\subsubsection{Objective Function}

Our objective function consists of two parts. The first part is the mean squared error (MSE) over multiple forward periods, which aims to minimize the prediction error. The second part is the contrastive learning loss, which guides the model to mine hidden factors that are both effective and comprehensive. The overall objective function is:
\begin{gather}
    \mathcal{L}_{\text{mse}} = \frac{1}{N \cdot L} \sum_{l=1}^{L} \sum_{i=1}^{N} (\hat{y}^{(i,l)} - y^{(i,l)})^{2} \\
    \mathcal{L} = \mathcal{L}_{\text{mse}} + \gamma  \mathcal{L}_{\text{CL}}
\end{gather}
where \( L \) is the number of forward prediction periods, \( y^{(i,l)} \) is the true return of the \( i \)-th stock at the \( l \)-th forward period, and \( \gamma \) is a hyperparameter that balances the contributions of the mean squared error loss and the contrastive learning loss.

\begin{table*}[ht]\fontsize{8pt}{10.5pt}\selectfont
    \centering
    \begin{tabular}{ll|cc|cc|cc|cc}
        \hline
        \multirow{2}{*}{Methods}                         & \multirow{2}{*}{Description}     & \multicolumn{2}{c|}{$\Delta t=1$} & \multicolumn{2}{c|}{$\Delta t=5$} & \multicolumn{2}{c|}{$\Delta t=10$} & \multicolumn{2}{c}{$\Delta t=20$}                                                                         \\
                                                         &                                  & IC                                & ICIR                              & IC                                 & ICIR                              & IC              & ICIR            & IC              & ICIR            \\ \hline
        \textbf{MLP}                                     & Multi-layer   perceptron model   & 0.0579                            & 0.6032                            & 0.0650                             & 0.7043                            & 0.0674          & 0.7155          & 0.0716          & 0.8029          \\
        \textbf{GRU} \cite{cho2014learning}              & RNN model based on   GRU         & 0.0637                            & 0.7045                            & 0.0813                             & 0.9277                            & 0.0829          & 0.9378          & 0.0861          & 1.0083          \\
        \textbf{TCN} \cite{bai2018empirical}             & Temporal   convolutional network & 0.0596                            & 0.6271                            & 0.0719                             & 0.8018                            & 0.0704          & 0.7880          & 0.0688          & 0.7986          \\
        \textbf{Transformer} \cite{vaswani2017attention} & Time-series   Transformer        & 0.0617                            & 0.6764                            & 0.0748                             & 0.8292                            & 0.0739          & 0.8523          & 0.0723          & 0.9102          \\
        \textbf{ALSTM} \cite{qin2017dual}                & Attention-based LSTM             & 0.0646                            & 0.7320                            & 0.0813                             & 0.9597                            & 0.0818          & 0.9687          & 0.0815          & 1.0091          \\
        \textbf{SFM}  \cite{zhang2017stock}              & Discrete Fourier transform       & 0.0621                            & 0.6758                            & 0.0789                             & 0.8425                            & 0.0821          & 0.8624          & 0.0853          & 0.9354          \\
        \textbf{GAT}  \cite{velivckovic2017graph}        & Graph attention   network        & 0.0538                            & 0.5197                            & 0.0667                             & 0.6743                            & 0.0678          & 0.7093          & 0.0669          & 0.7650          \\
        \textbf{HIST}  \cite{xu2021hist}                 & Mining hidden concepts           & 0.0571                            & 0.6059                            & 0.0705                             & 0.7954                            & 0.0703          & 0.8036          & 0.0622          & 0.7656          \\
        \textbf{HyperGCN} \cite{bai2021hypergraph}       & Hypergraph convolutional         & 0.0554                            & 0.5899                            & 0.0665                             & 0.7359                            & 0.0688          & 0.7502          & 0.0708          & 0.7552          \\
        \textbf{STHAN-SR}  \cite{sawhney2021stock}       & Spatio-temporal  hypergraph      & 0.0597                            & 0.6500                            & 0.0765                             & 0.9162                            & 0.0791          & 0.9724          & 0.0812          & 1.0201          \\
        \textbf{FactorVAE}  \cite{duan2022factorvae}     & A factor model based   on VAE    & 0.0646                            & 0.6699                            & 0.0807                             & 0.8680                            & 0.0756          & 0.7496          & 0.0722          & 0.6526          \\
        \textbf{CI-STHPAN}  \cite{xia2024ci}             & Spatial-temporal   pre-training  & 0.0554                            & 0.5460                            & 0.0707                             & 0.7570                            & 0.0727          & 0.7703          & 0.0711          & 0.7123          \\ \hline
        \textbf{FactorGCL}                               & Our proposed model               & \textbf{0.0684}                   & \textbf{0.7487}                   & \textbf{0.0885}                    & \textbf{0.9787}                   & \textbf{0.0915} & \textbf{1.0327} & \textbf{0.0929} & \textbf{1.0350} \\ \hline
    \end{tabular}
    \caption{The stock returns prediction performance of all compared methods on the test dataset; the higher, the better.}
    \label{tab:metric}
\end{table*}

\section{Experiments}

In this section, we present a series of experiments to demonstrate the effectiveness of our proposed method in real-world stock markets. Our discussion is structured around the following key research questions:

\begin{itemize}
    \item \textbf{RQ1:} How does our method compare to existing stock trend prediction methods in terms of performance?
    \item \textbf{RQ2:} What impact does each module in our model have on its overall performance?

    \item \textbf{RQ3:} How does varying the number of hidden factors affect the model's performance?
          % \item \textbf{RQ3:} What market characteristics are reflected by the hidden factors identified by our model?

    \item \textbf{RQ4:} Can our method achieve higher investment profits in simulated investment scenarios?

\end{itemize}

\subsection{Experiment Settings}

% \subsubsection{Dataset}

We conduct our experiments on the China A-shares market, utilizing a dataset spanning from 01/01/2014 to 06/30/2023. This dataset includes 5028 stocks, excluding suspended or otherwise abnormal stocks, and is constructed from a sequence of market data comprising day-level price-volume data (\textit{high}, \textit{open}, \textit{low}, \textit{close}, \textit{volume-weighted average price}, and \textit{trading volume}). In detail, the cross-sectional standardizated future stock returns with multiple periods ($\Delta t=1, 5, 10, 20$) are used as labels, and the future return is calculated by the formula $y_{t} = \frac{\text{price}_{t+\Delta t+ 1} - \text{price}_{t+1}}{\text{price}_{t+1}}$, where $\text{price}_{t}$ is the volume-weighted average price at trading day $t$. The length of historical squence data $x$ is $T=60$, and the length of future data $x'$ is $T'=20$. We adopt secondary industry factors as the prior factors (83 industries, if a stock belongs to the industry, the value of corresponding factor exposure is 1, otherwise 0).

We follow the temporal order to split the dataset into training set, validation set and test set, where the time length is \textit{5~years} : \textit{1~year} : \textit{2~years}, and adopt a rolling method for training and testing. The overall test period is from 01/01/2020 to 06/30/2023. The other details about the experiment are provided in the supplementary material\footnote{\url{https://tinyurl.com/bdhffkch}}.
% https://drive.google.com/file/d/1dObXIQMwg3-caQcQlg-gW6v_m1S8BOiH/view

\subsection{Main Results}
In the experiment, we compare our proposed model with competitive baselines on the stock trend prediction task. In order to evaluate the performance of the compared methods, we adopt the information coefficient (IC) as metric, which are the widely-used evaluation metrics in finance. Besides, we also report the information ratio of information coefficient (ICIR) to evaluate the stability of prediction.

Table~\ref{tab:metric} summarizes the performances of all compared methods on the test dataset. Our method achieves the highest IC and ICIR among all the compared methods. From the experimental results, we have the following observations:

\begin{itemize}
    \item FactorGCL can achieve better results than existing stock trend prediction methods like ALSTM, SFM and STHAN-SR, which illustrates the effectiveness of the proposed method for stock trend prediction.
    \item Moreover, compared with some baselines which can also extract the hidden relations from market data, like HIST, FactorVAE and CI-STHPAN, our model can mine the hidden factors more effectively and comprehensively, to achieve higher prediction performance.
\end{itemize}

\subsection{Ablation Study}

To verify the effectiveness of different modules in our framework, we build four variants of proposed model by removing the prior beta module, hidden beta module, individual alpha module, and contrastive learning loss, respectively. Table~\ref{tab:ablation} lists the IC of these variants on the test dataset. The results show that each module in our model contributes to the overall performance of our model.

\begin{table}[h]\fontsize{9pt}{12pt}\selectfont
    \centering
    \begin{tabular}{c|cccc}
        \hline
                               & $IC_{\Delta t=1}$ & $IC_{\Delta t=5}$ & $IC_{\Delta t=10}$ & $IC_{\Delta t=20}$ \\ \hline
        -\textit{wo Prior}     & 0.0654            & 0.0842            & 0.0904             & 0.0906             \\
        -\textit{wo Hidden}    & 0.0649            & 0.0819            & 0.0828             & 0.0851             \\
        -\textit{wo Alpha\&CL} & 0.0618            & 0.0808            & 0.0857             & 0.0856             \\
        -\textit{wo CL}        & 0.0661            & 0.0856            & 0.0899             & 0.0912             \\
        \hline
        FactorGCL              & \textbf{0.0684}   & \textbf{0.0885}   & \textbf{0.0915}    & \textbf{0.0929}    \\ \hline
    \end{tabular}
    \caption{The ablation study results on the test dataset}
    \label{tab:ablation}
\end{table}

\begin{figure}[t]
    \begin{centering}
        \includegraphics[width=0.90\columnwidth]{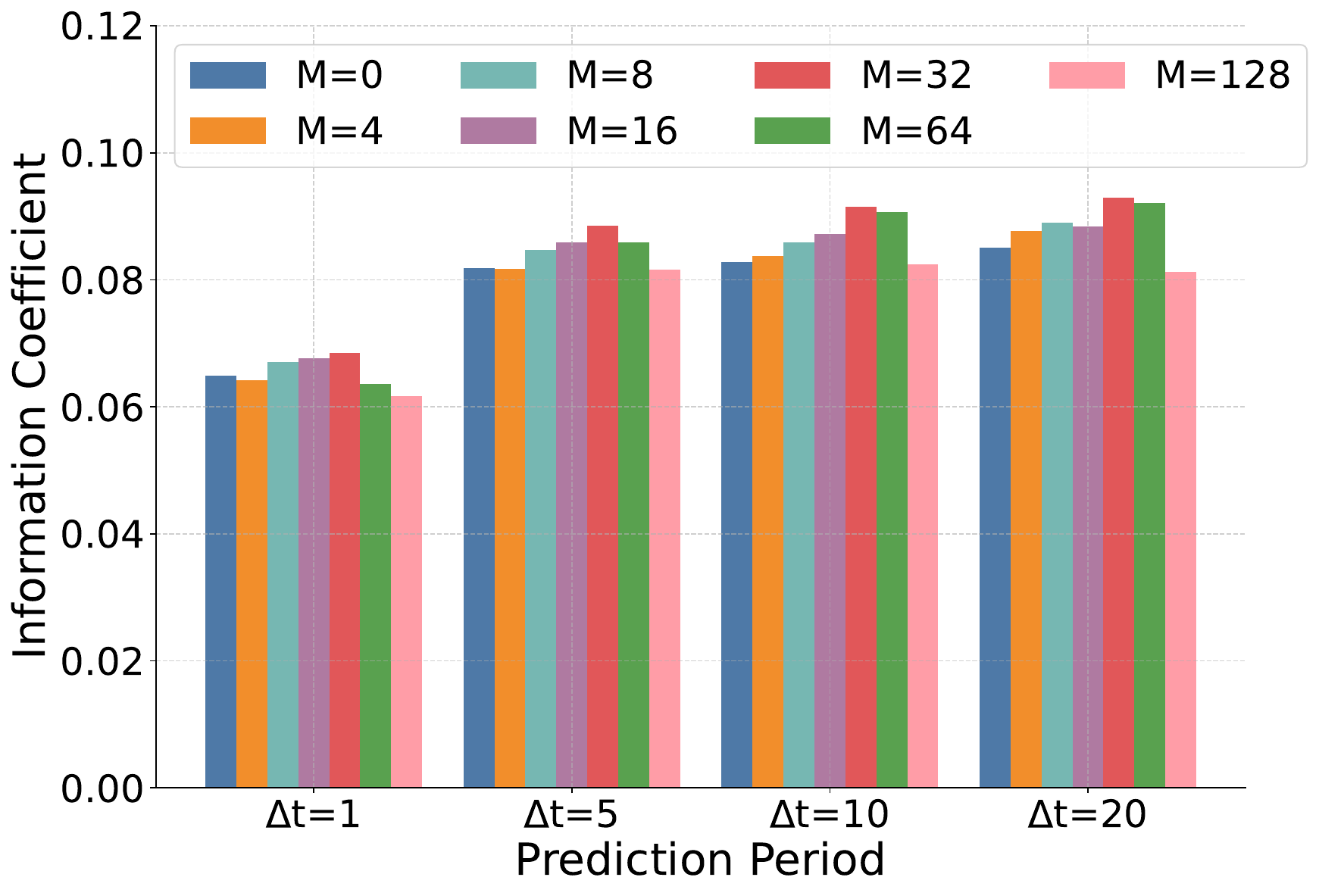}
        \caption{The performances of FactorGCL with different numbers of hidden factors.}
        \label{fig:hidden}
    \end{centering}
\end{figure}

We also conducted an experiment to assess the impact of varying the number of hidden factors \( M \) on the model's performance. In this experiment, we adjusted the number of hidden factors produced by the hidden beta module while keeping all other components of the model constant. The results, shown in Figure~\ref{fig:hidden}, demonstrate an improvement in model performance as the number of hidden factors increases, indicating that hidden factors play a crucial role in enhancing the model's predictive power.
However, beyond a certain point, further increases in hidden factors lead to a decline in performance, suggesting that an excessive number of hidden factors can cause overfitting and degrade the model's efficacy, which is consistent with our intuition.

\begin{table}[b]\fontsize{9pt}{10pt}\selectfont
    \centering
    \begin{tabular}{c|ccc|ccc}
        \hline
        \multirow{2}{*}{Methods} & \multicolumn{3}{c|}{CSI300} & \multicolumn{3}{c}{CSI500}                                                                     \\
                                 & AR                          & IR                         & RoMaD          & AR             & IR             & RoMaD          \\ \hline
        MLP                      & 0.034                       & 0.957                      & 0.180          & -0.001         & -0.015         & -0.003         \\
        GRU                      & 0.083                       & 2.281                      & 1.071          & 0.043          & 1.137          & 0.401          \\
        TCN                      & 0.042                       & 1.121                      & 0.418          & 0.010          & 0.261          & 0.067          \\
        Transformer              & 0.058                       & 1.675                      & 0.742          & 0.040          & 1.082          & 0.396          \\
        ALSTM                    & 0.078                       & 2.248                      & 0.891          & 0.046          & 1.323          & 0.413          \\
        SFM                      & 0.095                       & 2.913                      & 1.122          & 0.025          & 0.743          & 0.185          \\
        GAT                      & 0.004                       & 0.100                      & 0.037          & -0.020         & -0.515         & -0.117         \\
        HIST                     & 0.048                       & 1.369                      & 0.592          & -0.022         & -0.620         & -0.172         \\
        HyperGCN                 & 0.015                       & 0.454                      & 0.107          & 0.007          & 0.200          & 0.069          \\
        STHAN-SR                 & 0.047                       & 1.376                      & 0.544          & 0.033          & 0.906          & 0.165          \\

        FactorVAE                & 0.092                       & 2.446                      & 0.740          & 0.050          & 1.271          & 0.411          \\

        CI-STHPAN                & 0.044                       & 1.271                      & 0.251          & 0.028          & 0.797          & 0.252          \\
        \hline
        FactorGCL                & \textbf{0.169}              & \textbf{4.208}             & \textbf{2.978} & \textbf{0.087} & \textbf{2.092} & \textbf{0.737} \\ \hline
    \end{tabular}
    \caption{The investment simulation results on CSI300 and CSI500 stocks; the higher, the better.}
    \label{tab:backtest}
\end{table}

\subsection{Investment Simulation}

To further evaluate the profitability of our method in the real stock market, we conduct an investment simulation. Specifically, we adopt a simple stock selection strategy, referred to as the \textit{TopK} strategy. This strategy involves investing in the \textit{TopK} stocks with the highest predicted scores each trading day and selling them after holding for \(\Delta t\) days, where \(\Delta t\) represents the model's prediction period. In the simulation backtest, we select stocks from the CSI 300 and CSI 500 indexes, which are representative of large-cap and mid to small-cap stocks, respectively, providing a balanced and comprehensive evaluation of our investment strategy across different market segments. We use the equally weighted CSI300 and CSI500 portfolio as the benchmark, and set $\textit{TopK}=30$, $\Delta t=10$, and the transaction cost to 0.3\%.

\begin{figure}[t]
    \begin{centering}
        \includegraphics[width=1.0\columnwidth]{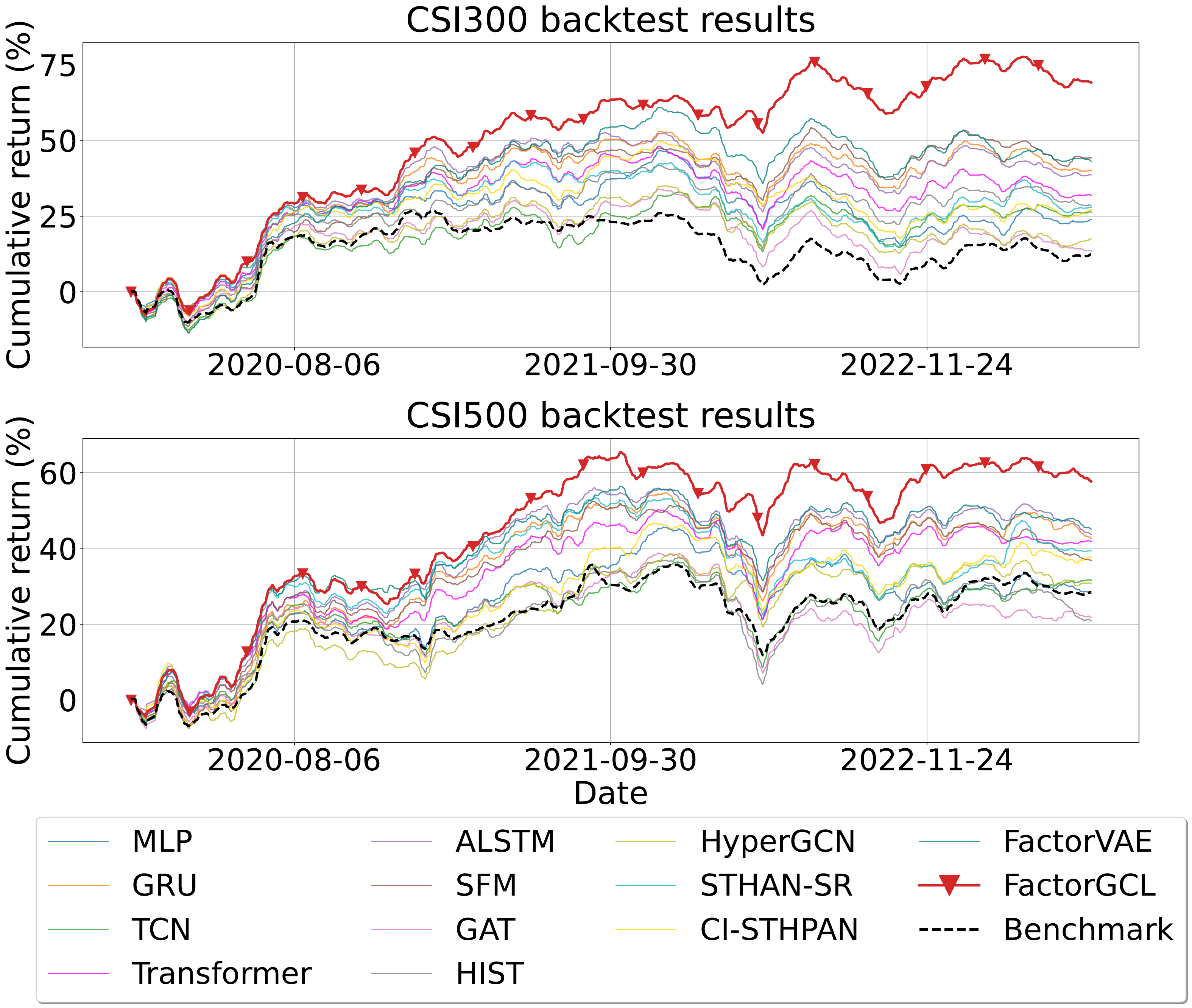}
        \caption{The results of investment simulation.}
        \label{fig:backtest}
    \end{centering}
\end{figure}

We present the cumulative return (CR) curves of all compared methods in Figure~\ref{fig:backtest} and report the annualized return (AR), information ratio (IR), and return over maximum drawdown (RoMaD) of cumulative excess return (CER) relative to the benchmark in Table~\ref{tab:backtest}, and the meaning of these metrics can be found in the supplementary material. The investment simulation results show that our method achieves the best performance across all metrics, indicating that our model can achieve profitable investments in the real market.

\section{Conclusion}

In this paper, we propose a novel hypergraph-based factor model with temporal residual contrastive learning (FactorGCL), which leverages both human-designed prior factors and data-driven hidden factors to predict stock returns. Specifically, our model follows a cascading residual hypergraph architecture, in which the hidden factors are extracted from the residual information after removing the prior factor information. To enhance the effectiveness and comprehensiveness of the hidden factors, we design a temporal residual contrastive learning method that contrasts stock-specific residuals embeddings over different time periods. Our extensive experiments demonstrate that our proposed FactorGCL outperforms existing state-of-the-art baselines in terms of predictive accuracy and investment profitability, and the ablation study further verifies our method can effectively extract hidden factors to improve the model's performance. In the future, we plan to apply the hypergraph-based factor model to risk factor mining and portfolio optimization.

\clearpage

\bibliographystyle{plain}

\begin{thebibliography}{40}
\providecommand{\natexlab}[1]{#1}

\bibitem[{Almeida and Freire(2023)}]{almeida2023nonlinear}
Almeida, C.; and Freire, G. 2023.
\newblock Which (Nonlinear) Factor Models?
\newblock \emph{Available at SSRN 4421179}.

\bibitem[{Bai, Kolter, and Koltun(2018)}]{bai2018empirical}
Bai, S.; Kolter, J.~Z.; and Koltun, V. 2018.
\newblock An empirical evaluation of generic convolutional and recurrent
  networks for sequence modeling. arXiv.
\newblock \emph{arXiv preprint arXiv:1803.01271}, 10.

\bibitem[{Bai, Zhang, and Torr(2021)}]{bai2021hypergraph}
Bai, S.; Zhang, F.; and Torr, P.~H. 2021.
\newblock Hypergraph convolution and hypergraph attention.
\newblock \emph{Pattern Recognition}, 110: 107637.

\bibitem[{Bansal and Yaron(2004)}]{bansal2004risks}
Bansal, R.; and Yaron, A. 2004.
\newblock Risks for the long run: A potential resolution of asset pricing
  puzzles.
\newblock \emph{The journal of Finance}, 59(4): 1481--1509.

\bibitem[{Caron et~al.(2020)Caron, Misra, Mairal, Goyal, Bojanowski, and
  Joulin}]{caron2020unsupervised}
Caron, M.; Misra, I.; Mairal, J.; Goyal, P.; Bojanowski, P.; and Joulin, A.
  2020.
\newblock Unsupervised learning of visual features by contrasting cluster
  assignments.
\newblock \emph{Advances in neural information processing systems}, 33:
  9912--9924.

\bibitem[{Chen and He(2021)}]{chen2021exploring}
Chen, X.; and He, K. 2021.
\newblock Exploring simple siamese representation learning.
\newblock In \emph{Proceedings of the IEEE/CVF conference on computer vision
  and pattern recognition}, 15750--15758.

\bibitem[{Cho et~al.(2014)Cho, Van~Merri{\"e}nboer, Gulcehre, Bahdanau,
  Bougares, Schwenk, and Bengio}]{cho2014learning}
Cho, K.; Van~Merri{\"e}nboer, B.; Gulcehre, C.; Bahdanau, D.; Bougares, F.;
  Schwenk, H.; and Bengio, Y. 2014.
\newblock Learning phrase representations using RNN encoder-decoder for
  statistical machine translation.
\newblock \emph{arXiv preprint arXiv:1406.1078}.

\bibitem[{Cui et~al.(2023)Cui, Li, Zhang, Guan, and Wang}]{cui2023temporal}
Cui, C.; Li, X.; Zhang, C.; Guan, W.; and Wang, M. 2023.
\newblock Temporal-relational hypergraph tri-attention networks for stock trend
  prediction.
\newblock \emph{Pattern Recognition}, 143: 109759.

\bibitem[{Daniel, Hirshleifer, and Sun(2020)}]{daniel2020short}
Daniel, K.; Hirshleifer, D.; and Sun, L. 2020.
\newblock Short-and long-horizon behavioral factors.
\newblock \emph{The review of financial studies}, 33(4): 1673--1736.

\bibitem[{Duan et~al.(2022)Duan, Wang, Zhang, and Li}]{duan2022factorvae}
Duan, Y.; Wang, L.; Zhang, Q.; and Li, J. 2022.
\newblock Factorvae: A probabilistic dynamic factor model based on variational
  autoencoder for predicting cross-sectional stock returns.
\newblock In \emph{Proceedings of the AAAI Conference on Artificial
  Intelligence}, volume~36, 4468--4476.

\bibitem[{Eugene and French(1992)}]{eugene1992cross}
Eugene, F.; and French, K. 1992.
\newblock The cross-section of expected stock returns.
\newblock \emph{Journal of Finance}, 47(2): 427--465.

\bibitem[{Fama and French(2021)}]{fama2021multifactor}
Fama, E.~F.; and French, K.~R. 2021.
\newblock \emph{Multifactor explanations of asset pricing anomalies}.
\newblock University of Chicago Press.

\bibitem[{Feng et~al.(2019)Feng, You, Zhang, Ji, and Gao}]{feng2019hypergraph}
Feng, Y.; You, H.; Zhang, Z.; Ji, R.; and Gao, Y. 2019.
\newblock Hypergraph neural networks.
\newblock In \emph{Proceedings of the AAAI conference on artificial
  intelligence}, volume~33, 3558--3565.

\bibitem[{Gu, Kelly, and Xiu(2021)}]{gu2021autoencoder}
Gu, S.; Kelly, B.; and Xiu, D. 2021.
\newblock Autoencoder asset pricing models.
\newblock \emph{Journal of Econometrics}, 222(1): 429--450.

\bibitem[{Han et~al.(2023)Han, Xie, Chen, and Xu}]{han2023stock}
Han, H.; Xie, L.; Chen, S.; and Xu, H. 2023.
\newblock Stock trend prediction based on industry relationships driven
  hypergraph attention networks.
\newblock \emph{Applied Intelligence}, 53(23): 29448--29464.

\bibitem[{He and Krishnamurthy(2013)}]{he2013intermediary}
He, Z.; and Krishnamurthy, A. 2013.
\newblock Intermediary asset pricing.
\newblock \emph{American Economic Review}, 103(2): 732--70.

\bibitem[{Hou et~al.(2021)Hou, Xu, Liu, Liu, Bian, Wu, Li, Chen, and
  Liu}]{hou2021stock}
Hou, M.; Xu, C.; Liu, Y.; Liu, W.; Bian, J.; Wu, L.; Li, Z.; Chen, E.; and Liu,
  T.-Y. 2021.
\newblock Stock trend prediction with multi-granularity data: A contrastive
  learning approach with adaptive fusion.
\newblock In \emph{Proceedings of the 30th ACM International Conference on
  Information \& Knowledge Management}, 700--709.

\bibitem[{Kelly, Pruitt, and Su(2019)}]{kelly2019characteristics}
Kelly, B.~T.; Pruitt, S.; and Su, Y. 2019.
\newblock Characteristics are covariances: A unified model of risk and return.
\newblock \emph{Journal of Financial Economics}, 134(3): 501--524.

\bibitem[{Levin(1995)}]{levin1995stock}
Levin, A. 1995.
\newblock Stock selection via nonlinear multi-factor models.
\newblock \emph{Advances in Neural Information Processing Systems}, 8.

\bibitem[{Li et~al.(2020)Li, Zhou, Xiong, and Hoi}]{li2020prototypical}
Li, J.; Zhou, P.; Xiong, C.; and Hoi, S.~C. 2020.
\newblock Prototypical contrastive learning of unsupervised representations.
\newblock \emph{arXiv preprint arXiv:2005.04966}.

\bibitem[{Li et~al.(2022)Li, Cui, Cao, Du, and Zhang}]{li2022hypergraph}
Li, X.; Cui, C.; Cao, D.; Du, J.; and Zhang, C. 2022.
\newblock Hypergraph-based reinforcement learning for stock portfolio
  selection.
\newblock In \emph{ICASSP 2022-2022 IEEE International Conference on Acoustics,
  Speech and Signal Processing (ICASSP)}, 4028--4032. IEEE.

\bibitem[{Li et~al.(2024)Li, Li, Shi, Xu, Du, Tan, Huang, and
  Lin}]{li2024alphafin}
Li, X.; Li, Z.; Shi, C.; Xu, Y.; Du, Q.; Tan, M.; Huang, J.; and Lin, W. 2024.
\newblock AlphaFin: Benchmarking Financial Analysis with Retrieval-Augmented
  Stock-Chain Framework.
\newblock \emph{arXiv preprint arXiv:2403.12582}.

\bibitem[{Lin et~al.(2022)Lin, Liu, Zhou, Hu, Wang, Zhao, Zheng, Lin, Xing, and
  Liang}]{lin2022prototypical}
Lin, S.; Liu, C.; Zhou, P.; Hu, Z.-Y.; Wang, S.; Zhao, R.; Zheng, Y.; Lin, L.;
  Xing, E.; and Liang, X. 2022.
\newblock Prototypical graph contrastive learning.
\newblock \emph{IEEE transactions on neural networks and learning systems},
  35(2): 2747--2758.

\bibitem[{Lintner(1975)}]{lintner1975valuation}
Lintner, J. 1975.
\newblock The valuation of risk assets and the selection of risky investments
  in stock portfolios and capital budgets.
\newblock In \emph{Stochastic optimization models in finance}, 131--155.
  Elsevier.

\bibitem[{Liu et~al.(2022)Liu, Zheng, Zhang, Chen, Peng, and
  Pan}]{liu2022towards}
Liu, Y.; Zheng, Y.; Zhang, D.; Chen, H.; Peng, H.; and Pan, S. 2022.
\newblock Towards unsupervised deep graph structure learning.
\newblock In \emph{Proceedings of the ACM Web Conference 2022}, 1392--1403.

\bibitem[{Oord, Li, and Vinyals(2018)}]{oord2018representation}
Oord, A. v.~d.; Li, Y.; and Vinyals, O. 2018.
\newblock Representation learning with contrastive predictive coding.
\newblock \emph{arXiv preprint arXiv:1807.03748}.

\bibitem[{Qin et~al.(2017)Qin, Song, Chen, Cheng, Jiang, and
  Cottrell}]{qin2017dual}
Qin, Y.; Song, D.; Chen, H.; Cheng, W.; Jiang, G.; and Cottrell, G. 2017.
\newblock A dual-stage attention-based recurrent neural network for time series
  prediction.
\newblock \emph{arXiv preprint arXiv:1704.02971}.

\bibitem[{Sawhney et~al.(2021)Sawhney, Agarwal, Wadhwa, Derr, and
  Shah}]{sawhney2021stock}
Sawhney, R.; Agarwal, S.; Wadhwa, A.; Derr, T.; and Shah, R.~R. 2021.
\newblock Stock selection via spatiotemporal hypergraph attention network: A
  learning to rank approach.
\newblock In \emph{Proceedings of the AAAI Conference on Artificial
  Intelligence}, volume~35, 497--504.

\bibitem[{Sawhney et~al.(2020)Sawhney, Agarwal, Wadhwa, and
  Shah}]{sawhney2020spatiotemporal}
Sawhney, R.; Agarwal, S.; Wadhwa, A.; and Shah, R.~R. 2020.
\newblock Spatiotemporal hypergraph convolution network for stock movement
  forecasting.
\newblock In \emph{2020 IEEE International Conference on Data Mining (ICDM)},
  482--491. IEEE.

\bibitem[{Sharpe(1964)}]{sharpe1964capital}
Sharpe, W.~F. 1964.
\newblock Capital asset prices: A theory of market equilibrium under conditions
  of risk.
\newblock \emph{The journal of finance}, 19(3): 425--442.

\bibitem[{Su et~al.(2024)Su, Wang, Qin, and Chen}]{su2024attention}
Su, H.; Wang, X.; Qin, Y.; and Chen, Q. 2024.
\newblock Attention based adaptive spatial--temporal hypergraph convolutional
  networks for stock price trend prediction.
\newblock \emph{Expert Systems with Applications}, 238: 121899.

\bibitem[{Treynor(1961)}]{treynor1961toward}
Treynor, J.~L. 1961.
\newblock \emph{Toward a theory of market value of risky assets}.

\bibitem[{Uddin and Yu(2020)}]{uddin2020latent}
Uddin, A.; and Yu, D. 2020.
\newblock Latent factor model for asset pricing.
\newblock \emph{Journal of Behavioral and Experimental Finance}, 27: 100353.

\bibitem[{Vaswani(2017)}]{vaswani2017attention}
Vaswani, A. 2017.
\newblock Attention is all you need.
\newblock \emph{arXiv preprint arXiv:1706.03762}.

\bibitem[{Veli{\v{c}}kovi{\'c} et~al.(2017)Veli{\v{c}}kovi{\'c}, Cucurull,
  Casanova, Romero, Lio, and Bengio}]{velivckovic2017graph}
Veli{\v{c}}kovi{\'c}, P.; Cucurull, G.; Casanova, A.; Romero, A.; Lio, P.; and
  Bengio, Y. 2017.
\newblock Graph attention networks.
\newblock \emph{arXiv preprint arXiv:1710.10903}.

\bibitem[{Xia et~al.(2024)Xia, Ao, Li, Liu, Liu, Ye, and Chai}]{xia2024ci}
Xia, H.; Ao, H.; Li, L.; Liu, Y.; Liu, S.; Ye, G.; and Chai, H. 2024.
\newblock CI-STHPAN: Pre-trained Attention Network for Stock Selection with
  Channel-Independent Spatio-Temporal Hypergraph.
\newblock In \emph{Proceedings of the AAAI Conference on Artificial
  Intelligence}, volume~38, 9187--9195.

\bibitem[{Xu et~al.(2021)Xu, Liu, Wang, Xia, Bian, Yin, and Liu}]{xu2021hist}
Xu, W.; Liu, W.; Wang, L.; Xia, Y.; Bian, J.; Yin, J.; and Liu, T.-Y. 2021.
\newblock Hist: A graph-based framework for stock trend forecasting via mining
  concept-oriented shared information.
\newblock \emph{arXiv preprint arXiv:2110.13716}.

\bibitem[{Zhang, Aggarwal, and Qi(2017)}]{zhang2017stock}
Zhang, L.; Aggarwal, C.; and Qi, G.-J. 2017.
\newblock Stock price prediction via discovering multi-frequency trading
  patterns.
\newblock In \emph{Proceedings of the 23rd ACM SIGKDD international conference
  on knowledge discovery and data mining}, 2141--2149.

\bibitem[{Zhou, Huang, and Sch{\"o}lkopf(2006)}]{zhou2006learning}
Zhou, D.; Huang, J.; and Sch{\"o}lkopf, B. 2006.
\newblock Learning with hypergraphs: Clustering, classification, and embedding.
\newblock \emph{Advances in neural information processing systems}, 19.

\bibitem[{Zou et~al.(2022)Zou, Cao, Liu, Lin, Abbasnejad, and
  Shi}]{zou2022astock}
Zou, J.; Cao, H.; Liu, L.; Lin, Y.; Abbasnejad, E.; and Shi, J.~Q. 2022.
\newblock Astock: A new dataset and automated stock trading based on
  stock-specific news analyzing model.
\newblock \emph{arXiv preprint arXiv:2206.06606}.

\end{thebibliography}

\end{document}